%% file: sample-sigconf.tex
\DocumentMetadata{}
\documentclass[manuscript]{acmart} 

\usepackage{multirow}
\begin{document}
\title{Electricity Consumption of Ethereum and Filecoin: Advances in Models and Estimates}
\subtitle{Pre-print Version}

\author{Elitsa Pankovska}
\affiliation{%
  \institution{Open Universiteit}
  \city{Heerlen}
  \country{Netherlands}
}
\email{elitsa.pankovska@gmail.com}

\author{Ashish Rajendra Sai}
\affiliation{%
  \institution{Open Universiteit}
  \city{Heerlen}
  \country{Netherlands}
}
\affiliation{%
  \institution{Maastricht University}
  \city{Maastricht}
  \country{Netherlands}
}
\email{ashish.sai@ou.nl}

\author{Harald Vranken}
\affiliation{%
  \institution{Open Universiteit}
  \city{Heerlen}
  \country{Netherlands}
}
\affiliation{%
  \institution{Radboud University}
  \city{Nijmegen}
  \country{Netherlands}
  }
\email{harald.vranken@ou.nl}

\author{Alan Ransil}
\affiliation{%
  \institution{Protocol Labs}
  \city{San Francisco}
  \state{California}
  \country{United States}
  }
\email{aransil@alumni.stanford.edu}

\renewcommand{\shortauthors}{E. Pankovska et al.}

\begin{abstract}
The high electricity consumption of cryptocurrencies that rely on proof-of-work (PoW) consensus algorithms has raised serious environmental concerns due to its association with carbon emissions and strain on energy grids. There has been significant research into estimating the electricity consumption of PoW-based cryptocurrencies and developing alternatives to PoW.

In this article, we introduce refined models to estimate the electricity consumption of two prominent alternatives: Ethereum, now utilizing proof-of-stake (PoS), and Filecoin, which employs proof-of-spacetime (PoSt). Ethereum stands as a leading blockchain platform for crafting decentralized applications, whereas Filecoin is recognized as the world's foremost decentralized data storage network.

Prior studies for modeling electricity consumption have been criticized for methodological flaws and shortcomings, low-quality data, and unvalidated assumptions. We improve on this in several ways: we obtain more novel, validated data from the systems in question, extract information from existing data and research, and we improve transparency and reproducibility by clearly explaining and documenting the used methodology and explicitly stating unavoidable limitations and assumptions made. When comparing the current, most prominent models for Ethereum and Filecoin to our refined models, we find that given the wide error margins of both the refined models and the ones introduced in prior literature, the resulting average estimates are to a large extent in line with each other.
%We prove certain assumptions in the Filecoin model, and we combine the strengths of separate Ethereum models and data to obtain an improved energy estimate, which takes into account additional information, such as the hosting distribution.

\end{abstract}



\maketitle

\input{samplebody-conf}

\bibliographystyle{ACM-Reference-Format}
\bibliography{sample-bibliography} 

\end{document}

%% file: samplebody-conf.tex
\section{Introduction}
While there is no question that Proof-of-Work (PoW)-based cryptocurrencies consume large amounts of electricity, claims that alternative mechanisms solve this problem need more rigorous scientific proof \cite{sai2022promoting}. However, estimating the electricity consumption of cryptocurrencies comes with multiple challenges. Due to the decentralized manner of operation, the hardware, and the efficiency of the network participants' operations vary greatly. Also, different consensus mechanisms require different types of resources, e.g., PoW requires computing power, while Filecoin's Proof-of-Spacetime (PoSt) mainly focuses on data storage. As a result, different electricity consumption patterns and trends follow.

There is already a significant amount of research in the field of modeling cryptocurrencies' electricity consumption \cite{sai2022promoting,vranken2017sustainability,sedlmeir2020energy}. However, the majority of these studies exist independently of each other or rely on unchallenged assumptions made in previous studies, highlighting the lack of appropriate scrutiny in publishing academic literature. Furthermore, the most common problem is the lack of rigour in the majority of published work \cite{sai2022promoting,lei2021best}. In this study, we want to advance the field by demonstrating how to design more reliable models by collecting novel data, basing model choices on the recommendation of the literature \cite{lei2021best, sai2022promoting}, and documenting the used methodology in a more traceable and verifiable manner.

% We structure our research around the following key research question:

% How to identify the limitations of existing cryptocurrency energy consumption estimation models and improve those by utilizing current research and data and working on reproducibility and transparency regarding unavoidable limitations?

For this purpose, we apply the code of conduct introduced by \cite{sai2022promoting} to analyze the weaknesses of existing models and propose standardized best practices for future research in the field. We focus on two cryptocurrencies - Ethereum and Filecoin. Ethereum is a prime example of a blockchain-based platform for building decentralized applications and recently replaced PoW with PoS (proof-of-stake), mainly to reduce electricity consumption \cite{kapengut2023event}. However, models estimating the electricity use of PoS-based Ethereum are still in their early stages. Filecoin is the world’s largest decentralized data storage network \cite{filecoin}. It applies proof-of-spacetime (PoSt), a novel consensus mechanism allowing participants to cryptographically prove they are storing data, and introduced provisions to incentivize green mining. Additionally, a basic model was already developed by the Filecoin Green team and since we have access to the Filecoin ecosystem, we believe we can improve the quality and reliability of the model. The paper makes the following contributions:

\begin{enumerate}
    \item We assess the strengths and weaknesses of existing models for estimating the electricity consumption of both Ethereum and  Filecoin.
    \item We develop an improved model for estimating the electricity consumption of PoS-based Ethereum, by analyzing and combining existing models and data about the network's size and different types of participants.
    \item We provide an improved model for estimating the electricity consumption of PoSt-based Filecoin, by validating assumptions made for Filecoin, collecting new data, and combining these with existing data.
    \item We show and explain the divergence between the existing models and estimates and our improved ones for both Filecoin and Ethereum.
\end{enumerate}
\section{Background}
This section provides a brief overview of the different approaches to modeling the electricity consumption of blockchain systems and the associated challenges. Additionally, we present the current research done in the subdomain of modeling Ethereum and Filecoin.

\subsection{Modeling Electricity Consumption}
We conducted a literature review using the following keywords: energy/electricity consumption/use, combined with one of the following: blockchain, cryptocurrencies, and distributed systems. In the context of cryptocurrencies, the terms energy consumption and electricity consumption are often used interchangeably, although electricity consumption is the more precise term, that we also use in this paper.

Overall, there is primarily research in the area of PoW-based cryptocurrencies, specifically Bitcoin. This is understandable given the high electricity demand of PoW and Bitcoin's financial value. Due to the lack of substantial work in the subdomain of modeling the electricity consumption of alternative consensus algorithms that aim at reducing electricity consumption, we focus our study on such algorithms in particular. We focus on Ethereum and Filecoin for the reasons explained in Section 1.

There are two main approaches to estimating the energy consumption of a decentralized network: bottom-up and top-down. In a bottom-up approach, exact electricity measurements are taken for all pieces of hardware that are considered present in the network and then the total network consumption is calculated by taking the hardware distribution in the network \cite{sai2022promoting}, which is often estimated via surveys. The major difficulty however is determining the hardware distribution in the network. This cannot be measured exactly, and estimates are used that rely on assumptions that cannot be fully validated. Moreover, experiments with cryptocurrencies with very high financial value, such as Ethereum and Bitcoin, are often limited due to a lack of sufficient budget. Furthermore, for Ethereum, there are discrepancies among data sources concerning the network's size and the composition of its nodes.

In a top-down approach, on the other hand, the model relies on high-level technical, economic, or social variables, usually due to the lack of sufficient empirical data \cite{sai2022promoting}. Most existing electricity models use this approach, including Shi et al.~ \cite{shi2023confronting}, who estimate the current and projected electricity consumption of the PoW and PoS-based Bitcoin and Ethereum blockchains by using variables such as hash rate, electricity price, market price for the PoW electricity estimation and the break-even point of staking rewards and costs for PoS-based Ethereum. %Such a top-down approach is for instance applied to PoS-based Ethereum in the experimental study conducted by the CCRI\cite{ccri2022}, which is discussed further in section 2.2.

Lei et al.~\cite{lei2021best} further split down the top-down approach into 4: classical top-down, economic, hybrid top-down, and extrapolation based on direct measurements. They suggest best practices for modeling the electricity consumption of blockchains and suggest that when possible, a bottom-up approach is recommended. In reality, most models are a hybrid version of the different approaches and while we also strive to bring our estimates as close to the truth as possible, we realize that analyzing distributed ledger technology (DLT) systems comes with certain limitations.

In the next two subsections, we delve into the research on electricity consumption modeling for PoS-based Ethereum and Filecoin and briefly explain Filecoin's network operations.

\subsection{Ethereum}
There are multiple studies, both scientific and from the industry, estimating the electricity consumption of Ethereum after the transition to PoS. Platt et al.~\cite{9741872} offer a model suitable for various PoS-based blockchains. While it can be useful for general comparison, the model results in at least 15 times broader margins, 3 times higher lower bound, and 15 times higher upper bound than alternative ones, which is due to inaccuracies about the network's size.

A drawback of recent studies is unrealistic assumptions regarding hardware. De Vries \cite{de2023cryptocurrencies} establishes a range for electricity consumption. They base the lower bound on the assumption that a 4GB Raspberry Pi meets the minimum hardware requirements, while the upper bound corresponds to the energy usage of an enterprise server. Platt et al.~\cite{9741872} use the same hardware setups and the general specifications for their expected power consumption at near full load. However, these hardware specifications do not adhere to the requirements for running a full Ethereum node. While some sources suggest that a node can be run even from a Raspberry Pi, the requirements for running a validator on top of it include a minimum of 16GB RAM. On the other hand, a desktop computer with 16/32GB RAM is sufficient to run a validator \cite{ethnode}, so the upper bound used in these studies is unnecessarily powerful, especially when running at full load is assumed.

A major drawback of the existing models by both Platt et al.~\cite{9741872} and Shi et al.~\cite{shi2023confronting} is assuming that each validator runs on a separate node. While that is theoretically possible, it does not hold for Ethereum currently. Running a node essentially means verifying the validity of every block and keeping the chain up-to-date on a piece of hardware without staking any ETH or adding new blocks \cite{ethnode}. Validators, on the other hand, stake ETH and add new blocks and currently, the system allows multiple validators to connect to one node and run on a single machine, with evidence suggesting that from a certain point onwards, adding more validators does not increase hardware resource usage significantly \cite{sutton2023}. There are over 700k validators in the network\cite{beaconchain}, but less than 13k nodes\cite{miganodesN}, out of which less than 7k have validators\cite{miganodesV}.

In terms of exact electricity measurements, none of the above-mentioned models make a distinction between the consumption of validators and simple nodes. The most reliable measurements are done by the CCRI \cite{ccri2022}, who have come up with 3 different hardware setups that all adhere to the minimum hardware requirements and run all possible combinations of execution and consensus clients on these setups. A significant downside is that these nodes do not run any validators due to the associated costs. These experiments are used as a backbone of the estimations done by both \cite{ccri2022} and \cite{cnsi}, however, the assumed hardware distribution is not based on any empirical evidence or literature findings.

While Ibanez et al.~\cite{ibanez2023energy} revisit the model developed by Platt et al., the updated estimates are taken directly from CCRI's measurements, combined with the number of nodes as reported on Ethernodes, which are lower than that of both Etherscan \cite{etherscan} and MonitorEth \cite{miganodesN}. This results in possibly unrealistically low estimates.

Also, to the best of our knowledge, none of the existing models take into account whether the nodes are hosted on a cloud, which changes their electricity consumption patterns significantly, due to the more efficient hardware and lower power usage effectiveness (PUE) of cloud providers. Multiple data sources suggest that nearly 50\% of the Ethereum 2.0 nodes use cloud services \cite{miganodesN,ethernodes-hosting}, the majority using AWS \cite{ethernodes-hosting2}.

All the Ethereum electricity estimation models we reviewed omit the PUE factor. In line with the recommendations of Sai et al.~\cite{sai2022promoting} and Lei et al.~\cite{lei2021best}, we consider that the PUE is a significant aspect of modeling Ethereum's electricity consumption, especially for nodes hosted via cloud services.

\subsection{Filecoin}
%In this section, we first briefly explain how Filecoin operates, and next present the current energy model, developed by the Filecoin Green Team.

Filecoin is the world's largest decentralized data storage network \cite{filecoin}, in which data is onboarded in a one-time setup process called sealing and then stored over time. Storage providers earn rewards by validating transactions and storing clients' data. Network participants operate via minerIDs, but bigger storage providers in the system may use multiple minerIDs for reasons such as distributed risk or having separate investors/clients, which will be discussed further in Section \ref{sec:methodology}. In the following, we consider storage providers (SPs) as being all individuals/companies that may operate as a group or on their own via one or more minerIDs.

SPs in the network submit proofs every 24 hours to attest they are still storing the data they have sealed. The two processes that consume the most electricity in the system are sealing and storing data, as shown in the experiments conducted by Pankovska et al.~\cite{pankovska2023determining}. In Filecoin's current energy model \cite{ransil2021}, two further aspects are not considered: the electricity used to generate these daily proofs and the electricity consumed to produce and validate blocks. The reason for this is that compared to the electricity used for sealing and storing data, these two aspects do not contribute to the total electricity use of the system significantly\cite{ransil2021}.

Filecoin’s current model is an electricity estimation model based on dimensional analysis with several limitations such as a lack of data quality control and traceability. It is based on survey data with 15 complete responses, 2 interviews with SPs, hardware specifications of hard disk drives, and the thermal design power of hardware used for sealing sectors \cite{ransil2021}. The energy consumption of the network is calculated via the following formula:

\begin{equation}
  P = ( A \cdot SR + B \cdot Cap) \cdot PUE
\end{equation}
Where:

\begin{itemize}
    \item P (Electrical Power, Watts) is the total electricity use.
    \item A (Sealing Electricity Constant, Wh/byte) is the electricity required per byte to seal a sector.
    \item SR (Sealing Rate, bytes/hour) is the rate at which new data is being sealed, determined from on-chain proofs.
    \item B (Storage Power Constant, W/byte) is the electrical power per byte required to store data over time.
    \item Cap (Raw Capacity, bytes) is the amount of data stored, determined from on-chain proofs.
    \item PUE (Power Usage Effectiveness, dimensionless) is the Power Usage Effectiveness, which is the ratio of total electrical power consumed to that consumed by useful IT processes such as sealing and storage.
\end{itemize}
The Sealing Rate and Raw Capacity can be determined using on-chain proofs, so the fundamental aspect of modeling Filecoin's electricity consumption is to
determine the upper bound, lower bound, and estimated
values of A, B, and PUE.

To calculate these values, the Filecoin Green team asked all survey participants about (1) the storage capacity they have added to the network in the past 28 days, (2) the average power use of their sealing hardware (in kWh/day), and (3) the power consumption and capacity of a storage rack in their system. Then, A and B can be calculated in the following way:

\begin{equation}
  A = \frac{P_{seal} \cdot 28 \cdot 1000}{C_{28days}} 
\end{equation}
\begin{equation}
  B = \frac{P_{storage} \cdot 1000}{C_{rack}} 
\end{equation}

Where:

\begin{itemize}
    \item $P_{seal}$ (kWh/day) is the daily average power use of the sealing hardware
    \item $C_{28days}$ (bytes) is the sealed storage capacity in the last 28 days.
    \item $P_{storage}$ (kW) is the power consumption of a storage rack
    \item $C_{rack}$ (bytes) is the capacity of a storage rack
\end{itemize}

For each available data point, an estimate for the constants A and B is calculated, outliers are removed, and a lower-, upper bound and an average estimate is chosen by picking one out of the calculated constants. The weakness of this method is that a single value, which is not necessarily the median, nor the mean is picked to estimate the consumption of the whole system, relying solely on domain knowledge and individual examples, rather than proving a statistical significance or a linear relationship between the independent variables (sealed and stored data) and the dependent one (electricity consumption). In this study, we aim to analyze whether such relationships exist and build models relying on more data from various sources. The chosen PUE values are taken from a study by Masanet et al.~\cite{masanet2020recalibrating} and average data center industry estimates by the Uptime Institute \cite{uptime2021}.

\section{Methodology and Results}
\label{sec:methodology}

In this section, we explain the methodology used to collect and analyze data and model the electricity consumption of two cryptocurrencies: Ethereum and Filecoin. For both systems, we begin the process of model development by identifying room for improvement of the models found in previous literature. We collect additional data, re-analyze, and compare the existing data sources that give us information about the network and its participants. We systematically combine all the insights we have gathered, following the guidelines set by Sai et al. \cite{sai2022promoting} and come up with lower- and upper bounds and average estimates for the electricity consumption of Ethereum and Filecoin.

\subsection{Ethereum}

We model Ethereum's electricity consumption by identifying the number and type of network participants and estimating their individual power demand. The main aspects we need to consider are the network's size, the hosting distribution of network participants, the hardware they use, and its electricity consumption \cite{ccri2022, 9741872, de2023cryptocurrencies}, as well as the PUE \cite{sai2022promoting}.

There are 4 main sources of information regarding the node count in the Ethereum network: Etherscan \cite{etherscan}, Ethernodes \cite{ethernodes-hosting}, Nodewatch \cite{nodewatch}, and MonitorEth \cite{miganodesN}. The first two are stated on Ethereum's official website but provide neither methodology nor source code for the way they source their data. The methodology of Nodewatch is not explained explicitly, however, they do provide their source code. The most transparent source of information is MonithorEth, which is essentially a fork of Nodewatch with available code and a research paper analyzing the network and its participants \cite{cortes2021discovering}. To our knowledge, all of these sources work more or less in a similar way: essentially, they gather information via a network crawler - an Eth2 node that requests the Ethereum Node Records (ENRs) of other nodes in the network.

Unfortunately, the node counts differ significantly between sources. We suspect that some of the sources might count unsynced nodes or only count validator nodes without explicitly stating so because half of the sources report a number of nodes around 6k while the others have a much higher number over 10k. Due to the more transparent methodology and the fact that MonithorEth is also used as a source of information by the research done by both the CCRI and the CBECI, we also decide to use it as a data source.

Moreover, an important aspect of the data presented by MonitorEth is the distinction between validator and non-validator nodes. At the time of research, there are 6066 validator nodes \cite{miganodesV} and 4710 non-validator ones \cite{miganodesN}. While all nodes in the network must have at least 2TB SSD of available storage, the RAM requirements for validators are 16GB, while 4GB is sufficient for running a simple node \cite{ethnode}. As a result, the lower-bound hardware setups differ depending on the node type.

\begin{table}
  \caption{Node Counts depending on type and hosting}
  \label{tab:ethcounts}
  \begin{tabular}{ccc}
    \toprule
    Node Type&Hosting&Count\\
    \midrule
    Validator & Cloud & 2182\\
    Validator & Other & 3884\\
    Non-validator & Cloud & 2867\\
    Non-validator & Other & 1843\\
    
  \bottomrule
\end{tabular}
\end{table}

Furthermore, the main difference we introduce as opposed to other models is that we use different estimates for the power demand of nodes/validators hosted on a cloud vs. home setups. Ethernodes, Nodewatch, and MonitorEth provide information about the hosting distribution in the network, with the number of nodes hosted on a cloud varying between 2205 and 4731. This probably stems from the difference in total node counts as well, however, their proportion also varies between 15\% and 58\%. Here, we again choose to trust MonithorEth, resulting in node counts per type and hosting as seen in Table \ref{tab:ethcounts}. Only Ethernodes provides a breakdown of the distribution of different hosting providers. More than half of the nodes hosted on a cloud use AWS, with all other providers having a comparatively negligible portion of 6.1\% or less \cite{ethernodes-hosting2}. Therefore, we use energy estimates of AWS EC2 instances, by relying on empirical experiments by \cite{ec2consumption}. These experiments are conducted with minimal amounts of storage, so we add 5W\footnote{5W was the highest consumption for a piece of SSD we could find and we aim for more conservative estimates, so likely the realistic consumption might be lower.} due to the 2TB SSD requirements.

There is no single right type of EC2 instance for running an Eth node, with various online sources suggesting different options: m5\cite{urbanek}, m5a\cite{awsm5a}, and t2\footnote{\url{https://docs.google.com/document/d/1ug-UruaXsghWy_0qvcUWOnJT9ltFho8rQxrIo7vv3Tk/edit}}. For running a validator, a minimum of 16GB is recommended, which corresponds to the xlarge size of all 3 types of EC2 instances. For running a simple node, large (8GB) is sufficient.

The source which suggests an m5.xlarge instance also shows the CPU load, which never reaches 100\% but is consistently over 50\%. For this reason, we use the electricity estimates for instances operating at full load, since the other option is 50\%. For the lower bound of the non-validator nodes, we chose instances with 8GB, while for validators we use the minimum of 16GB. Since previous work suggests that both types of nodes can work smoothly with 16GB of RAM \cite{urbanek, awsm5a, ethnode}, we keep this configuration for the average estimate and 32GB of RAM as the upper bound. Although even more powerful instances exist, they also come at a significantly higher cost, so we do not believe that network participants would opt for those due to financial reasons. Table \ref{tab:ethhardwarecloud} shows the resulting power demand for each of the bounds as well as the estimate, which we calculate by taking the average between the power demands of the instance types mentioned above (m5, m5a, t2) and adding 5W on top due to the power demand of the SSD.

\begin{table}
  \caption{Conceivable upper and lower bounds for the power demand of a
cloud validator/non-validator instance}
  \label{tab:ethhardwarecloud}
  \begin{tabular}{ccccc}
    \toprule
    Node Type&Config&Instance Type&Demand(W)\\
    \midrule
    Non-validator & Minimum & large & 21.4W\\
    Validator & Minimum & xlarge & 37.4W\\
    Both & Medium & xlarge & 37.4W\\
    Both & Maximum & 2xlarge & 70.5W\\
    
  \bottomrule
\end{tabular}
\end{table}

\begin{table*}
  \caption{Conceivable upper and lower bounds for the power demand of a
home setup validator/non-validator machine}
  \label{tab:ethhardware}
  \begin{tabular}{ccccc}
    \toprule
    Node Type&Config&Hardware Type&Exemplar&Demand(W)\\
    \midrule
    Non-validator & Minimum & Small single-board computer& Raspberry Pi & 6.4W\\
    Validator & Minimum & Pre-built desktop& Intel i5-1135G7 (CCRI Measurement) & 20W\\
    Both & Medium & Average estimate of multiple configs & CCRI Measurements & 62.44W\\
    Both & \multirow{2}{4em}{Maximum} & Custom-built desktop & AMD 3970X (CCRI Measurement) & \multirow{2}{2.5em}{150W}\\
    & & Enterprise server & Ethstaker Hardware Examples & \\
    
  \bottomrule
\end{tabular}
\end{table*}

As already discussed, some hardware setups used for the estimates of previous models do not adhere to the minimum requirements or are unnecessarily powerful. Multiple consensus and execution clients require as little as 4GB RAM for running a full Ethereum node \cite{ccri2022}. This, however, is not sufficient for running a validator, so we have two separate lower bounds for validators and simple nodes, similarly to the estimates for EC2 instances.

Multiple sources suggest that running a full node is possible even on a Raspberry Pi \cite{hardwareexamples, stevan}, which is considered to be one of the computers with the lowest power demand\footnote{We use the measurements by \href{https://www.pidramble.com/wiki/benchmarks/power-consumption}{Raspberry Pi Dramble}}. For the lower bound of a validator, we have chosen the same bound as the CCRI, which is a pre-built desktop computer with 16GB RAM.

We use the same average estimate as the CCRI, which is calculated by assuming the hardware distribution between 3 different hardware setups with 16, 64, and 256GB RAM, where 50\% of the nodes uses the middle setup and two others are used by 25\% each. This might be an overestimate of the real consumption because most network participants would not invest in unnecessarily powerful hardware. However, this is the only exact measurement we have access to, and even in it, the potential additional consumption due to adding multiple validators is not studied. Thus, we suspect that the two factors might cancel each other out to some extent.

For the upper bound, we consider two types of hardware: the custom-built desktop with 256GB RAM from CCRI's measurements and an enterprise server. Interestingly enough, both of those are reported to use about 150W\cite{ccri2022, hardwareexamples}. A summary of the power demand of each of the bounds can be seen in Table \ref{tab:ethhardware}.

The last aspect of the power demand of the network is the PUE. The PUE is the ratio of total electricity used to that consumed by IT infrastructure, the additional non-IT electricity consumers usually being cooling and lighting. While cloud providers such as Google report values as low as 1.1 \footnote{\url{https://www.google.com/about/datacenters/efficiency/}}, we decide to use a slightly more conservative value of 1.2 for all nodes hosted on a cloud, by relying on an AWS Blog Post \cite{awspue}.

There is generally less information about the PUE of a home setup, however, a study by Shehabi et al.~\cite{shehabi2016united} estimates a PUE of 2.0 for a closet and 2.5 for a room in 2014 and predicts values of 2.0 and 1.7 for 2020, respectively. Thus, we take the more conservative value of 2.0 for a closet setup and assume that would be the PUE of all nodes that are not hosted on a cloud.

Using the data as outlined above, the power consumption of the whole network can be calculated via the following formula:
\begin{equation}
  P = \sum_{j \epsilon \{v, n\}}\sum_{i \epsilon \{c,o\}}^{} (n_{ij} \cdot p_{ij}) \cdot PUE_i
\end{equation}

Where $v$ and $n$ correspond to the type of node: $v$ for a validator and $n$ for a normal node and $c$ and $o$ correspond to the hosting type: $c$ for cloud and $o$ for other. $n_{ij}$, $p_{ij}$ correspond to the number of nodes and the power demand for that type of node, respectively.

% In addition, the power consumption per transaction (in Wh/tx) is:

% \begin{equation}
%   P_t = \frac{P}{3600 \cdot l}
% \end{equation}

% Where l is the system throughput in transactions per second.\footnote{We multiply it by 3600 because an hour consists of 3600 seconds.} The average throughput over the last 6 months is 12.1, calculated via the public data on Blockchair.\footnote{\url{https://blockchair.com/ethereum}}

\subsection{Filecoin}

In this section, we explain how we collect and analyze more data about the Filecoin network and utilize it to improve Filecoin Green's current electricity estimation model.

Other than the data the Filecoin electricity model is currently based on, we make use of a few additional data sources: system metrics regarding the number and size of minerIDs, an energy validation process (EVP) during which SPs provide information about their electricity bills, etc., and more detailed interviews with various SPs in the network. The computing resources of Filecoin SPs are extremely similar to those of standard data centers \cite{filecoin}, which means we can rely on PUE data from both existing scientific research and industry examples.

\subsubsection{Number and size of minerIDs}
We analyze the current system and its composition by using the public Spacescope API \cite{spacescope}, via which we can fetch open network data. To date, more than 40\% of the minerIDs in the network provide 1PB of raw byte power (RBP) or less. However, this large group holds a negligible chunk of the RBP of the whole network, less than 4\%. We divide minerIDs into 3 groups: with less than 1PB, 1 to 50PB, and over 50PB. We did so after consulting the Filecoin Green team, who claim that 1PB serves more or less as a borderline, beyond which SPs have to switch to a more professional setup. The group of large SPs over 50PB are usually those who already have their own data center, whereas the mid-sized SPs traditionally rent out space in a co-located DC. Given that the electricity consumption of the network depends mainly on the amounts of data that minerIDs are sealing/storing, the most relevant group for our analysis is the mid-sized minerIDs.

\subsubsection{Reassessing initial survey data}
We reassessed the data from the initial survey conducted by the Filecoin Green team. Three of the authors independently apply a predetermined methodology for analyzing the survey answers \footnote{See \url{https://github.com/elip06/filecoin-electricity-model} for official methodology, survey data, and data analysis code} and then compare the results from the coding step to check agreement/disagreement and, lastly, calculate Cohen's kappa value, which shows a substantial agreement. In addition to the initial survey, we extracted 2 additional answers regarding the sealing energy and 7 additional ones for storage energy.

We derived the value $3.13e-08$ for the sealing constant A (see equation 2). We verified that there is a linear relation between $C_{28days}$ and $P_{seal}$. The linear correlation between these (after removing clear outliers that are suspected to have occurred due to methodological error) is 0.68, which indicates that these variables are moderately to highly correlated. The values for A range between $1.03e-08$ and $1.15e-07$. The p-value from OLS (Ordinary Least Squares) regression is 0.00014, which confirms that a linear model is indeed suitable.

Similarly, we plot $C_{rack}$ against $P_{rack}$ using the survey answers, and upon outlier removal, the variables are highly correlated with a correlation coefficient of 0.97, B ranging between $5.21e-13$ and $1.00e-11$. The resulting p-value for the linear model is $3.44e-11$, showing that the independent variable is highly significant and the resulting estimate for B is $5.86e-12$.

\subsubsection{Reassessing EVP data}
To further test and confirm the reliability of the survey results, we study the data from Filecoin's EVP. During the EVP, Filecoin SPs are given the opportunity to provide information about their operations, hardware setup, electricity, and water consumption. In return, they are awarded a sustainability tier, based on the information they have provided and an audit done by third-party validators. Not all of this data is publicly available, but we do have access to the monthly electricity consumption of the SPs, usually for a period between 9 months and a year, as well as the minerIDs which were operational at the corresponding time\footnote{see \url{https://www.greenscores.xyz/auditoutputs}}.

Using the API to the Filecoin energy dashboard\footnote{\url{https://filecoin-green.gitbook.io/filecoin-green-documentation/filecoin-green-api-docs/list-of-apis/energy-consumption-api}}, we fetch additional information about the amounts of data that were sealed and stored for a period of time of our choice. Our goal is to model Filecoin's electricity consumption using these sources of data namely - the data storage capacity being added (sealed) to the network and the total data storage capacity stored. There is sufficient data from 2 SPs, however, the data stored by one of them is not yet sealed and is therefore not recorded in Filecoin's dashboard, although it contributes largely to their electricity consumption. Consequently, we use the 11 months' worth of data provided by the other SP to perform linear regression with 2 independent variables: sealed data and stored data, and one dependent one: monthly electricity consumption. The resulting estimates are $1.64e-08$ for A and $3.21e-12$ for B, both being highly significant with p-values of $0.003133$ and $0.000302$, respectively. We use the SP's real reported PUE of 1.3.

\subsubsection{Interview Data}
Our last source of information is interviews with SPs in the network with various characteristics. By consulting Filecoin representatives and analyzing the data from the EVP, we have concluded that even large SPs usually operate via multiple smaller minerIDs. This means that even within this group of seemingly small providers, many are part of larger operations, which are usually associated with higher efficiency and lower PUE. This is the reason we focus our interviews on large SPs. Multiple large SPs during the interviews report that they have multiple individual clients, which delegate their right to manage their operations as a miner to larger entities. This means they have a say in the amount of data they seal and store on the network, but they do not own any hardware or perform any of the sealing/storing themselves.

We obtained data from 7 SPs with specific profiles. We conducted interviews with 5 SPs, while 2 SPs answered the interview questions by mail. A recent study\footnote{\url{https://observablehq.com/@jimpick/provider-quest-zoomable-sunburst?collection=@jimpick/provider-quest}}, the purpose of which is to map IPs and additional minerID information to locations, shows that the majority of minerIDs are located in Asia, North America, Europe, and Australia. We interviewed 1 SP from China, 2 from the US, 1 from Canada, 1 from Europe, and 2 from Australia and New Zealand, so we consider that as a good representation of the real system. Out of all 7, one SP only owns a single minerID under 1PB and should therefore be considered small, all others being mid- to large-sized.

During the interviews, SPs provide information in particular about their hardware, the power demand of both sealing and storage hardware, whether they rent out space in a co-located DC, their PUE, energy mix, number of minerIDs \footnote{See \url{https://github.com/elip06/filecoin-electricity-model/tree/main/SP_interviews} for the full question list, and the anonymized interview transcripts}. We collect PUE data from the SPs, since most of them rent out space in co-located DCs, which readily provide this information to their clients. Some SPs also provide exact hardware specifications and official electricity reports, which will be kept private due to security reasons.

Using the answers from the SPs, we calculate the power demand of their sealing (A) and storage (B) hardware and their PUE. Table \ref{tab:interviews} depicts the results from the interviews, as well as the estimates we have calculated using the survey data and the EVP. 

\subsubsection{PUE}
Additionally, we research the PUE values of DCs that offer space for rent internationally. While not all DCs provide information about their PUE, we found 10 that have locations in North America, Europe, Asia, and Australia with PUEs ranging between 1.2 and 1.79 with an average of 1.435. While we understand that PUE values can be dependent on SP location, we decided not to use the geographic distribution of nodes for the PUE calculation for two reasons: The location of nodes can only be determined by their IP address, which is not a reliable source due to the use of VPNs or minerIDs not wanting to release their location at all; Most SPs have provided very similar PUE values despite their very different locations/climates.

\begin{table}
  \caption{Results from interviews, EVP, and survey}
  \label{tab:interviews}
  \begin{tabular}{cccc}
    \toprule
    Source & A(Wh/byte) & B(W/byte) & PUE\\
    \midrule  
    Interview & 9.98e-09 & 1.13e-12 & 1.29\\
    Interview & 3.64e-08 & 2.99e-12 & 1.75\\
    Interview & 9.36e-09 & 1.04e-12 & 1.3\\
    Interview & 8.13e-09 & 9.22e-13 & 1.5\\
    Interview & - & - & 1.3\\
    Interview & 7.86e-09 & 6.22e-13 & 1.33\\
    Interview & - & 4.82e-12 & -\\
    EVP & 1.64e-08 & 3.21e-12 & - \\
    Survey & 3.13e-08 & 5.86e-12 & - \\  
  \bottomrule
\end{tabular}
\end{table}

\subsubsection{Results}
We calculated the average estimates for A, B, and PUE by combining all data and calculating weighted averages, where the weights are associated with the number of data points in the respective data sources. The lower- and upper bounds for both A, B, and PUE are the lowest/highest possible values which were not considered outliers out of all data sources, see a summary of the resulting model in Table \ref{tab:filmdel}.

\begin{table}
  \caption{Final Filecoin Model}
  \label{tab:filmdel}
  \begin{tabular}{cccc}
    \toprule
     & A(Wh/byte) & B(W/byte) & PUE\\
    \midrule  
    Lower & 7.86e-09(Interview) & 5.21e-13(Survey) & 1.2(Online)\\
    Average & 2.17e-08 & 4.16e-12 & 1.426\\
    Upper & 1.15e-07(Survey) & 1.00e-11(Survey) & 1.79(Online)\\
  \bottomrule
\end{tabular}
\end{table}

\section{Electricity Consumption Estimates}

\subsection{Ethereum}
Table \ref{tab:results} shows a comparison between the currently existing Ethereum electricity consumption models and our improved one and includes the lower- and upper bound of the electricity consumption, as well as an estimate that is considered to be the most realistic. Our predictions are closer to the ones done by the CCRI and Cambridge. It is worth noting, however, that while our average estimate is well within the error margins set by these two previous studies, the whole interval is shifted up. The lower bound we have set is 35\% higher than the average between CCRI's and Cambridge's lower bound, our average estimate is 17\% higher, and our upper bound is 12\% higher. Altogether, the lower power demand of the cloud setups is compensated by the addition of the PUE, resulting in higher estimates overall.

While the Digiconomist reports a nearly twice higher average estimate than ours, the methodology behind this estimate is not publicly available, so we cannot identify where this considerable difference comes from. The estimates from the further two studies, depicted in Table \ref{tab:results}, are, as already mentioned in Section 2.2, unrealistically high due to the wrong assumption that each validator runs on a separate machine.

Altogether, with this study, we show that multiple additional factors influence the electricity consumption of Ethereum than the ones considered in other studies.

\begin{table*}
  \caption{Energy Demand of Ethereum and Filecoin according to Existing Models and Our Improved Ones}
  \label{tab:results}
  \begin{tabular}{llccc}
    \toprule
    Ethereum Model & Publication Date & Lower Bound & Estimate & Upper Bound\\
    \midrule
    CCRI Crypto Sustainability Indices \cite{ccri2022} & Live (August 2023) & 260.35kW & 812.83kW & 1953.44kW\\
    Cambridge Ethereum Electricity Consumption Index \cite{cnsi} & Live (August 2023)& 256.69kW & 791.04kW & 1890.90kW\\
    Digiconomist \cite{digiconomist} & Live(August 2023) & - & 1717.25kW & - \\
    The Energy Footprint of Blockchain Consensus Mechanisms & September 2021 & 1010.6kW & - & 30887.5kW\\
    Beyond Proof-of-Work \cite{9741872}\\
    Confronting the Carbon-footprint Challenge of Blockchain \cite{shi2023confronting} & December 2021 & 4520.55kW & - & 35600.5kW\\
    \textbf{Our improved Ethereum Model} & \textbf{August 2023} & \textbf{350.5kW} & \textbf{941.79kW} & \textbf{2145.25kW} \\
    \\
    \midrule
    Filecoin Model&Publication Date&Lower Bound & Estimate & Upper Bound\\
    \midrule
    Filecoin Model \cite{fildashboard} & Live(Aug 31, 2023) & 14.454MW & 67.757MW & 212.909MW\\
    \textbf{Our improved Filecoin Model} & \textbf{Live(Aug 31, 2023)} & \textbf{9.39MW} & \textbf{79.086MW} & \textbf{257.917MW}\\
    
  \bottomrule
\end{tabular}
\end{table*}

We do not compare our results to Ethereum before switching to PoS, since we have not validated the methodology for estimating the electricity consumption of PoW-based Ethereum. However, even if we assume that the electricity estimates by both Digiconomist and Cambridge for PoW-based Ethereum are pessimistic to some extent, PoS-Ethereum's consumption is still several magnitudes lower, both as a total power demand of the network and on a per-transaction basis.
\subsection{Filecoin}
The average estimates in our improved Filecoin model are similar to the original ones, see Table \ref{tab:results}. Given the wide error margins, our average estimate is very much in line with the initial model, so although the final estimate is 17\% higher, this difference might not be considered significant. Our main contribution here is the scientific reasoning and the larger amount of data that support these values and that we validate the initial assumptions made regarding the linear relationship between electricity consumption and sealed/stored data. Also, the PUE values we utilize in the improved model are based primarily on empirical data collected from Filecoin SPs. We have set 35\% lower lower- and 21\% higher upper bounds than the original model, mainly because of considering fewer survey answers as outliers.

\section{Discussion}
\subsection{Interpretations}
The main goal of this study is to show the potential of improving existing electricity estimation models in the cryptocurrency domain by using additional data and more rigorous scientific research. The delivered results confirm that although accurate to some extent, existing estimates often rely on limited data and multiple assumptions about how the network is built and which factors contribute to its electricity consumption. We present ways in which such assumptions can be checked and conduct a more diligent scientific review of existing literature and data. Subsequently, we improve both models' methodology, reliability, and transparency.

We manage to identify the weaknesses of current models by studying the systems' composition and we build more trustworthy estimates by collecting additional empirical data about the system participants (in the Filecoin case) and making use of existing research from various sources (in Ethereum's case). We show that most existing Ethereum models underestimate the cryptocurrency's electricity consumption, in terms of lower-, upper bound, and average estimate. While the average estimate we have come up with for Filecoin is well within the original bounds, we believe a wider error margin is needed due to the uncertainty in the data the model is based on.

\subsection{Limitations}
Since we are working with decentralized systems, there is no single source of information that can be considered true for the whole network. This is the reason that we cannot measure the exact accuracy of our predictions. However, since we adhere to the best practices suggested by researchers in the field, we believe we deliver more rigorous and transparent electricity estimates.

Altogether, the majority of the information we use in modeling both cryptocurrencies' electricity consumption comes from non-scientific sources, which leaves room for inaccuracies and biased data. However, this is an unavoidable limitation when researching DLT systems. In this study, we compare and analyze data from various sources in order to minimize the effect of this limitation and come up with the most realistic estimates.

The main limitation of the Ethereum model is the unknown hardware distribution of the network, as well as the exact cost and electricity usage associated with running one or more validators on a single machine. This is the reason we opt for the electricity usage at near full load for each hardware setup in order to end up with more conservative estimates rather than significantly underestimating the electricity use of the system.

The small amount of data used for modeling Filecoin's electricity consumption can lead to inaccurate estimates. The reason for this is that the SPs from the EVP overlap to some extent with those from the interviews, meaning that in reality, we rely on a handful of SPs for our estimates for the whole network. We do, however, believe, that these SPs are good representatives of the system and since they all use similar hardware and report similar measurements, the network average will as a minimum fall within the same order of magnitude. Moreover, although not all of the SPs we interviewed shared a full list of their minerIDs, the total of the ones who did is 121, which hold over 90PB (0.7\%) of the RBP of the system. We interviewed SPs who actively take part in Filecoin Green's initiatives toward green mining. Most of them generate electricity from renewable sources onsite or purchase renewable energy certificates to compensate for their emissions. Also, given the small amount of survey data, outlier removal could not be based on statistical methods.

\section{Conclusion}
There has been significant interest in estimating various cryptocurrencies' electricity consumption, mainly due to their growth in value and high power demand (mostly in the case of PoW-based blockchains). While existing research focuses predominantly on Bitcoin, minimizing DLT systems' power demand by using alternative consensus algorithms, such as PoS, has gained a lot of attention lately. The overall opinion is that PoS is a much less electricity-consuming algorithm, although studies that deliver electricity estimates for PoS-based systems report highly varying numbers, putting into question how much exactly the electricity reduction is, compared to a PoW-based cryptocurrency. That is the case because estimating the electricity consumption of distributed systems comes with a lot of challenges due to the wide variety of participants' hardware and manner of operation.

The goal of this study is to focus on two case studies: Ethereum and Filecoin, analyze the existing models estimating their electricity consumption, and find and implement ways to improve them. We manage to validate the suitability of used modeling techniques and improve those using scarce data. For the Ethereum model, we choose hardware setups that adhere to the official hardware requirements and collect more reliable data regarding the size of the network and the different types of nodes. We expand on the existing knowledge and deliver more trustworthy estimates, relying on various data sources and information from actual network participants. We find that the original Filecoin model \cite{ransil2021} can benefit from wider error margins due to the data quality, but has an average estimate that is in line with our results as well. We find that both CCRI's \cite{ccri2022} and Cambridge's \cite{cnsi} Ethereum models likely underestimate the networks' electricity use, while the models by Platt et al.~\cite{9741872} and Shi et al.\cite{shi2023confronting} greatly overestimate it.

Future work should further study the reliability of the used sources. With sufficient resources, the impact of validator counts running on a single node on the electricity consumption of Ethereum should be empirically tested. Furthermore, the energy used for additional IT processes, such as networking and submitting daily proofs in the Filecoin network should be studied further in order to validate the assumption that sealing and storage are the main contributors to electricity usage. Via further data collection, the carbon footprint of both systems can also be estimated.
\appendix

\begin{acks}
We would like to thank the Filecoin Green team, including Marc Johnson for their valuable feedback and assistance in improving the model. We would also like to express our gratitude to the Filecoin storage providers we interviewed (DCENT, DSS, FilSwan, Holon, Mongo, PikNik, SXX) who provided us with invaluable insights into their operations.
\end{acks}